\documentclass[prl,twocolumn,superscriptaddress,notitlepage,longbibliography]{revtex4-2}
\usepackage{amsmath}
\usepackage{amssymb}

\usepackage[unicode=true,colorlinks=true,citecolor=blue,urlcolor=blue]{hyperref}

\usepackage{bm}
\usepackage{epsfig}
\usepackage{multirow}

\usepackage{graphicx}

\usepackage[normalem]{ulem}
\renewcommand {\Im}{\mathop\mathrm{Im}\nolimits}
\renewcommand {\Re}{\mathop\mathrm{Re}\nolimits}
\newcommand {\Tr}{\mathop\mathrm{Tr}\nolimits}

\renewcommand {\phi}{{\varphi}}
\newcommand {\rmi}{{\rm i}}

\newcommand {\e}{{\rm e}}

\begin{document}
\title{
{Multimer  states in multilevel waveguide QED
}
}

\author{Jiaming Shi}
\email{jiaming.shi@weizmann.ac.il}

\author{Alexander N. Poddubny}
\email{poddubny@weizmann.ac.il}
\affiliation{Department of  Physics of Complex Systems, Weizmann Institute of Science, Rehovot 7610001, Israel}

\begin{abstract}
{
We study theoretically the interplay of spontaneous emission and interactions for the multiple-excited quasistationary eigenstates in a finite periodic array of multilevel atoms coupled to the waveguide. We develop an analytical approach 
to calculate such eigenstates based on the subradiant dimer basis.
Our calculations reveal the peculiar multimerization effect driven by the anharmonicity of the atomic potential: while a general eigenstate is an entangled one, there exist eigenstates that are products of dimers, trimers, or tetramers, depending on the size of the system and the fill factor. At half-filling, these product states {acquire a periodic structure with all-to-all connections inside each multimer} and become the most subradiant ones.
}
\end{abstract}
\maketitle 

The interplay of interactions and dissipation in the distributed nonlinear system can give  rise to peculiar spatial patterns ranging from liquids \cite{Manneville2006,Cassani_2021} to nonlinear optics \cite{Arecchi1995,Zhang2015}.  Periodic pattern formation has also been studied in quantum systems \cite{Lonard2017,Li2017,Hertkorn2021}.
Nonlinear optimization for the states with minimal losses or maximal optical gain has also been recently proposed for analog computation, based on coupled laser networks~\cite{Tradonsky2019} or coupled optical parametric oscillators~\cite{McMahon_2016}.

\begin{figure}[b]
\centering\includegraphics[width=0.48\textwidth]{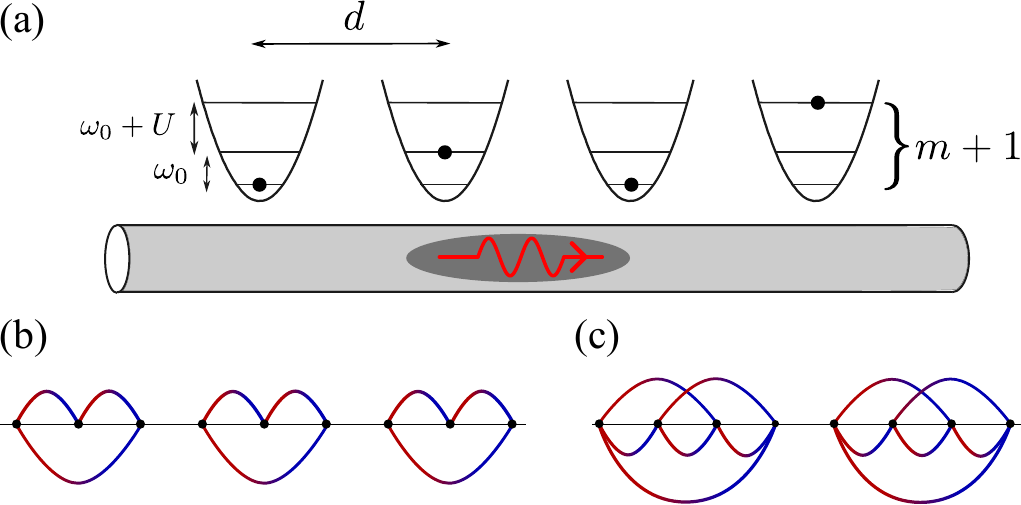}
\caption{(a)  Illustration of an array of multi-level atoms  coupled via photons in a waveguide. (b,c) Schematics of trimer and tetramer states in the array. Each red-blue line from site $m$ to site $n$ in  the diagrams corresponds to a subradiant dimer excitation $\sigma_n^\dag-\sigma_m^\dag$.}\label{fig:1}
\end{figure}

Here, we explore the spatial structure of the multiple-excited  states in the many-body quantum system, a finite periodic array of multilevel atoms, coupled to the waveguide, see Fig.~\ref{fig:1}. Such structures   have now become available based on the superconducting qubits, optical atoms, and trapped cold atoms, and the corresponding field of waveguide quantum electrodynamics (waveguide QED) has emerged~\cite{sheremet2021waveguide}. The system is inherently open because of the possibility of photon emission into the waveguide and features long-range photon-mediated couplings. The two-level atomic arrays are now relatively well studied both theoretically and experimentally. 
{In particular, for the chiral two-level atom arrays, peculiar steady-state multipartite entangled patterns have been theoretically predicted in Ref.~\cite{Pichler2015}.}
In non-chiral arrays, the transition from weak\cite{Molmer2019} to strong excitation regime has been studied in Ref.~\cite{Poshakinskiy2021dimer}, and the formation of periodic dimerized eigenstates has been predicted. Double-excited subradiant states have recently been demonstrated in a 4-qubit array~\cite{zanner2021coherent}.
However, much less is known about a more general type of  atoms with $m>1$ excited states, which correspond to actual superconducting transmon qubits~\cite{Krantz2019}. Another  relevant platform for the waveguide QED with multilevel atoms is a matter-wave QED platform~\cite{de_Vega_2008,Kwon2022}, where the role of photons is played by the unbound atomic states in the optical  lattice~\cite{kim2023super}.

Here, we show that the periodic pattern formation is a universal effect for  multilevel atoms coupled to the waveguide: while  most quasistationary  eigenstates of the atomic array are spatially entangled, there also exist several peculiar   multimerized product states:
\begin{multline}\label{eq:multimer}
    \psi_{1,2\ldots N}=\chi_{1,2\ldots p}\otimes
    \chi_{p+1,p+2\ldots 2p}\otimes\ldots
    \\\otimes \chi_{N-p+1,N-p+2\ldots N}
        \:.
\end{multline}
Here, the indices $1\ldots N$ label the atoms, the  array size $N$ is a product of two integers, $N=pq$, and $p$ is the size of the multimer. 
The spectra of eigenstates with the same fill factor are split into different branches due to the anharmonicity of the atomic potential. Importantly, for fill factors $f\le 1/2$, these multimer states are the most subradiant ones in their spectral branches.  Examples of trimer ($p=3$) and tetramer ($p=4$) product states  are illustrated  {by diagrams} in Fig.~\ref{fig:1}(b,c), where each line between two sites $n$ and $m$ corresponds to a subradiant dimer excitation of the atoms $n$ and $m$ with opposite phases $\sigma_n^\dag-\sigma_m^\dag$. One should multiply all such terms to obtain a multiple-excited subradiant eigenstate.
We will show, that in the case of  strong anharmonicity the number of lines per atom at half-filling condition $f=1/2$ is fixed to $m$, which restricts the states to the products of multimers.

\begin{figure*}[t]
\centering\includegraphics[width=0.99\textwidth]{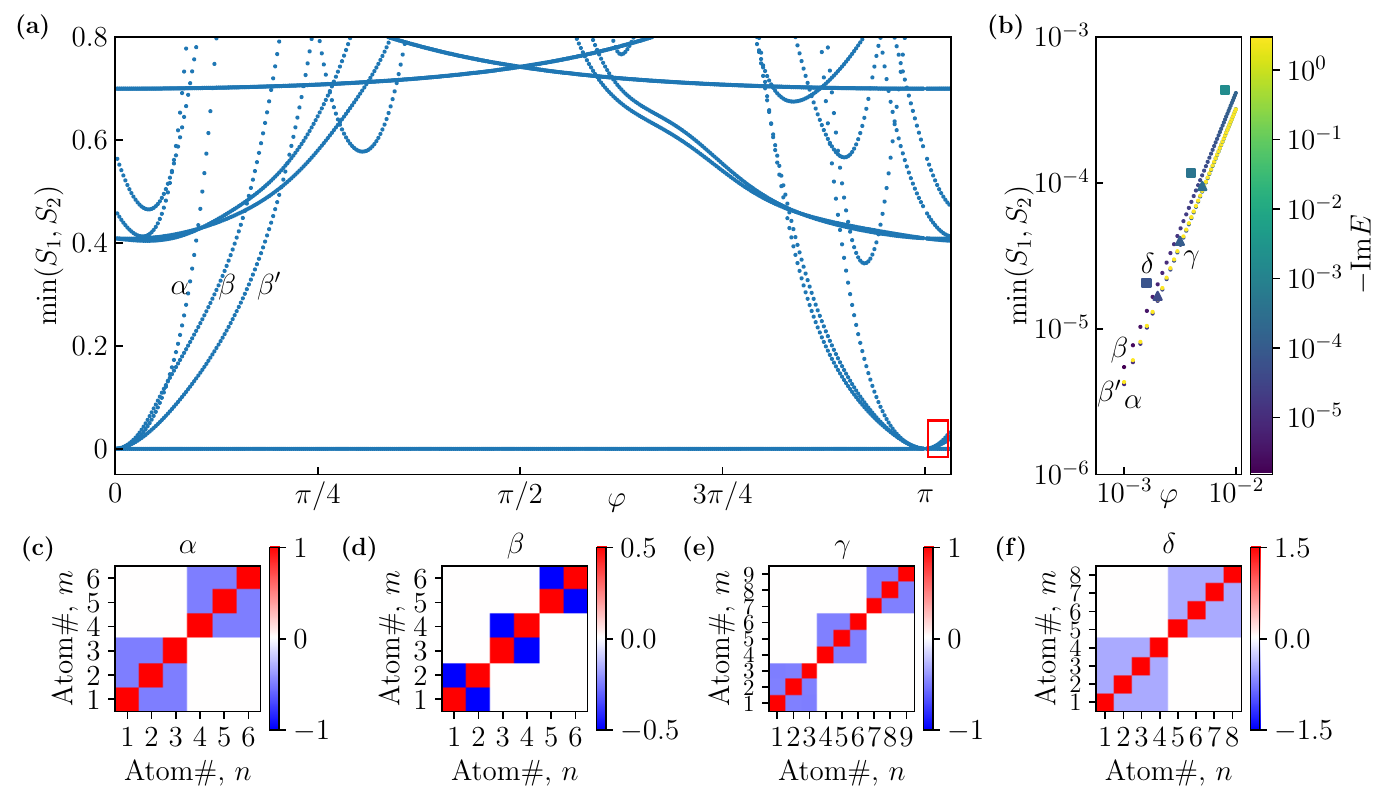}
\caption{(a) Dependence of the minimum value of entanglement entropy $S_1$ and $S_2$ for all the states in the 6-atom array with 3 levels per atom on the array period $\varphi$. The vertical axis is restricted to  $\text{min}(S_1,S_2) \leq 0.8$. Here $\alpha$ is the trimer state at half-filling, $f=1/2$. $\beta$ is the dimer state at  {$f=1/4$}. $\beta'$ is the particle-hole inversion of $\beta$. (b)  Zoomed-in  view of the red rectangle in (a). Here we add several data points of trimer state at $N=9, m=2, f=1/2$ as $\gamma$ and tetramer state at $N=8,m=3,f=1/2$ as $\delta$. (c),(d),(e),(f) The correlation $\langle \sigma_m^\dagger \sigma_n\rangle$ for the labeled states at $\varphi$ approaches 0. The calculations are performed for  $U/\gamma_{\rm 1D} = 5$.
}
\label{fig:entropy}
\end{figure*}
{\it Theoretical model.}  
{We now proceed to the rigorous description of the multimerization.} We consider  a basic waveguide QED setup with  $N$ multilevel atoms, periodically spaced near a waveguide and interacting via a waveguide mode. The system is schematically shown in Fig.~\ref{fig:1} and can be described by the following effective Hamiltonian~\cite{Caneva2015,sheremet2021waveguide}
\begin{equation}\label{eq:H}
	\begin{split}
	    H = \omega_0 &\sum_{n=1}^{N} \sigma_n^\dagger \sigma_n + \frac{U}{2}\sum_{n=1}^{N} \sigma_n^\dagger \sigma_n (\sigma_n^\dagger \sigma_n -1) \\
     -&\rmi \gamma_{\rm 1D} \sum_{m,n=1}^{N} \e^{\rmi \varphi |m-n|} \sigma_m^\dagger \sigma_n\:.
	\end{split}
\end{equation}
{We assume the Bose-Hubbard term $U$ is  nonzero but  much smaller than $\omega_0$. In this case, the usual Markovian and rotating-wave approximations underlying the effective Hamiltonian Eq.~\eqref{eq:H} still hold.} The raising operators $\sigma_i$ obey the usual  commutation relationship $[\sigma_i, \sigma_j] = \delta_{i,j}$ and $(\sigma_i^\dagger)^{m+1} = 0$, where $m+1$ is the number of levels {per atom}.  The phase is $\varphi = \omega_0 d/c$, where $d$ is the distance between neighboring qubits and the parameter $\gamma_{\rm 1D}$ is the radiative decay rate of a single atom into the waveguide.

{\it Multimer states.}
The two key parameters of our model are the relative anharmonicity strength $U/\gamma_{\rm 1D}$ and the phase $\varphi$, that controls the photon-mediated coupling between the atoms. 
For $\varphi=0,\pi\ldots$, the coupling is purely dissipative, and for $\varphi=\pi/2,3\pi/2\ldots$ it has an exchange character. Our goal is to analyze how the spatial profile of the multiple-excited eigenstates of Eq.~\eqref{eq:H} for larger anharmonicity depends on the coupling character. To this end, we  calculate the entanglement entropy $S$ in the  array of six 3-level atoms depending on $\varphi$. The array is divided into two subsystems $A$ and $B$ and the entanglement entropy is \cite{Eisert2010}
 $   S = - \Tr_A(\rho_A \log_2 \rho_A)$\:,
where $\rho_A = \Tr_B(\rho_{AB})$ is the reduced density matrix. We further define $S_{1,2}$ as the entanglement entropy with array $A$ containing the first three or two atoms, respectively.  Figure~\ref{fig:entropy} presents the minimal value $\text{min}(S_1, S_2)$ for each state depending on $\varphi$. The overall picture has a period of $\pi$~\cite{Ivchenko1994b}. The line at $S_{1,2}=0$, corresponds to the trivial fully excited state of the array. {Our main finding  is the presence of the states $\alpha,\beta,\beta'$ where $\text{min}(S_1, S_2)\to 0$ for $\varphi\to 0,\pi$ and quadratically increases
as $\varphi$ is detuned from $\varphi$ [see also zoomed in Fig.~\ref{fig:entropy}(b)].
} In order to understand  these states we plot in 
Fig.~\ref{fig:entropy}(c--d) their corresponding correlation functions $\langle \sigma_n^\dag \sigma_m\rangle.$ We find that  the state labelled as $\beta$ is the  dimerized state~\cite{Poshakinskiy2021dimer},
\begin{equation}
    (\sigma_1^\dagger-\sigma_2^\dagger)(\sigma_3^\dagger-\sigma_4^\dagger)(\sigma_5^\dagger-\sigma_6^\dagger) |0\rangle
    \label{eq:dimer}
\end{equation}
at $\varphi \to 0$ and state $\beta'$ is just the particle-hole inversion of state $\beta$. {One can replace the ground state $|0\rangle$ with the full excited state $|F\rangle$ and all the creation operators $\sigma_i^\dagger$ with the annihilation operators $\sigma_i$ in Eq.~\eqref{eq:dimer} to obtain the expression of state $\beta'$.} The state $\alpha$ is a  trimer state, at $\varphi\to 0$ {it is given by }
\begin{multline}\label{eq:trimer}
    (\sigma_1^\dagger-\sigma_2^\dagger)(\sigma_2^\dagger-\sigma_3^\dagger)(\sigma_3^\dagger-\sigma_1^\dagger)\times\\ (\sigma_4^\dagger-\sigma_5^\dagger)(\sigma_5^\dagger-\sigma_6^\dagger)(\sigma_6^\dagger-\sigma_4^\dagger) |0\rangle\:,
\end{multline}
in agreement with the general ansatz Eq.~\eqref{eq:multimer}. 
Inspired by this finding, we also numerically calculate the entanglement entropy of the most subradiant states at {half-filling} for  $N=9$, $m=2$ and $N=8$, $m=3$.  This yields two  more states $\gamma,\delta$ with $\min S\to 0$ at $\varphi \to 0$, see Fig.~\ref{fig:entropy}(b). Their correlations are plotted in Fig.~\ref{fig:entropy}(e) and Fig.~\ref{fig:entropy}(f). Here, the entanglement entropy of the states $\gamma$ and $\delta$ is calculated for subsystem $A$ containing  the first three and four atoms, respectively. Hence, the state $\gamma$ is a direct product of three trimers, as illustrated in Fig.~\ref{fig:1}(b), and state $\delta$ is a direct product of two tetramers [see Fig.~\ref{fig:1}(c)]. We will further term them a trimer state and a tetramer state.

\begin{figure}[t]
    \centering\includegraphics[width=0.49\textwidth]{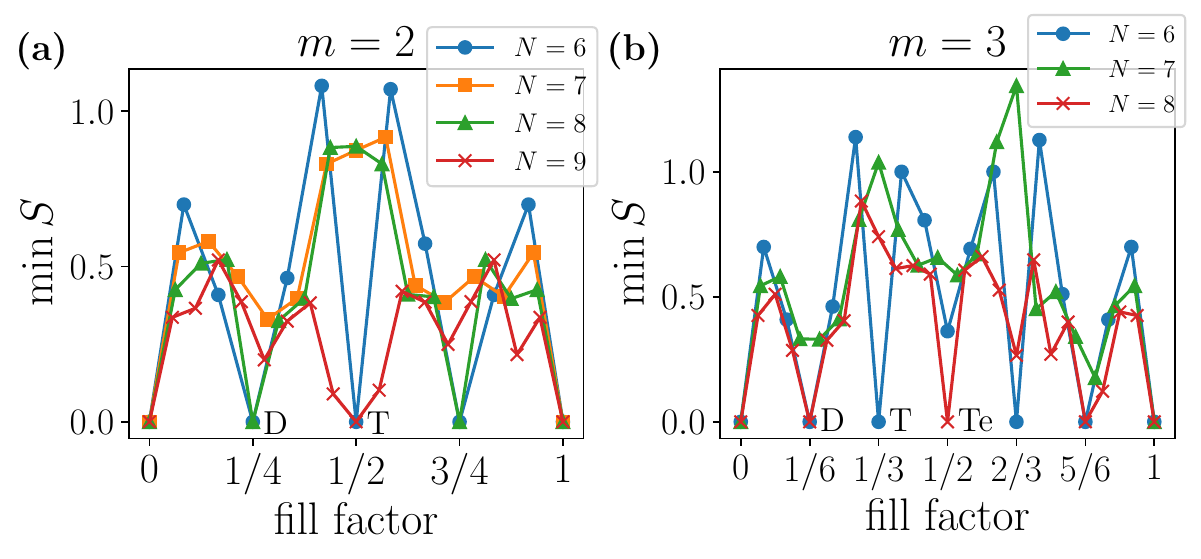}
\caption{Dependence of the minimum value of entanglement entropy on fill factor $f$ calculated at $\varphi = 0.001$,  $U/\gamma_{\rm 1D}=5$ for (a) $m=2$ and (b) $m=3$. The positions of dimer, trimer,  and tetramer states are marked with D, T and Te.}
\label{fig:minS-f}
\end{figure}

The presence of the multimer states with zero entanglement entropy, found in Fig.~\ref{fig:entropy}, turns out to be a general mesoscopic property of the finite system that is manifested for properly matching the number of atoms $N$, the number of excited levels per atom $m$ and the fill factor $f$. 
{Here we define the fill factor for each eigenstates as $f = k/(mN)$.
where $k$ is the number of excitations.
This is possible because  the Hamiltonian Eq.~\eqref{eq:H} commutes with the excitation number operator $\mathcal{N} = \sum_{n=1}^N \sigma_n^\dagger \sigma_n$.}

Figure~\ref{fig:minS-f} summarizes our calculations of  entanglement entropy for $N=6\ldots 9$ with $m=2$ [Fig.~\ref{fig:minS-f}(a)] and {$N=6\ldots 8$ with} $m=3$ [Fig.~\ref{fig:minS-f}(b)]. 
 For every eigenstate we calculated the entropy for all possible divisions of the array into  two parts with {at least}  two atoms in each part. Next,  for each value of the fill factor $f$ we present   the minimum value of $S$. The calculation in Fig.~\ref{fig:minS-f} shows sharp minima corresponding to the specific values of $f$ when the number of $N$ fits an integer multiple of dimers, trimers or tetramers.
In particular, when $m=2$, the dimer state and trimer state are allowed. For multiples of 3 and 2 like $N=6$, there is a dimer state at $f=1/4$, the particle-hole inversion of that dimer state at $f=3/4$ and a trimer state at $f=1/2$. For $N$ being just a multiple of 3 like $N=9$, there is only a trimer state at half-filling. For $N$ being just a multiple of 2 like $N=8$, there is only a dimer state at $f=1/4$ and its inversion. For $N=7$ neither dimer nor trimer states are possible. {We note that the dimer states at $f=1/4$ here can be considered as inherited from the array with two levels per atom at half-filling.}

When $m=3$, tetramer states are allowed. Next, for the multiples of 4 like $N=8$, there is a tetramer state at half-filling. Besides, for multiples of 2 like $N=6, 8$, there are dimer states at $f=1/6$ and their inversion at $f=5/6$. For multiples of 3 like $N=6$, there is a trimer state at $f=1/3$ and its inversion at $f=2/3$. For a prime number of $N=7$ no multimer states are possible.

As shown in Fig.~\ref{fig:entropy}(b) and Fig.~\ref{fig:minS-f}, the multimer states at $f \leq 1/2$ are subradiant. We now  present a perturbative approach to study the subradiant states in short-period arrays when $\varphi$ is close to 0.
Taylor expansion of the atom-photon interaction part of the Hamiltonian Eq.~\eqref{eq:H} results in
\begin{equation}\label{eq:perturbation}
	\begin{split}
	     - \rmi \sum_{m,n=1}^N \sigma_m^\dagger \sigma_n \e^{\rmi \varphi |m-n|} \approx -\rmi H_0 -\varphi H_1 -\frac{\rmi\varphi^2}{2} H_2\:,
	\end{split}
\end{equation}
where
\begin{align}\label{eq:perturbation2}
			H_0 &= \sum_{m,n=1}^N \sigma_m^\dagger \sigma_n,\quad 
		H_1 = -\sum_{m,n=1}^N \sigma_m^\dagger \sigma_n	|m-n|\:,\nonumber\\    
		H_2 &= -\sum_{m,n=1}^N \sigma_m^\dagger \sigma_n	|m-n|^2\:.
	\end{align}
We are looking for subradiant states, where the expectation value of $H_0$ is zero and the imaginary part of complex energies is proportional to $-\varphi^2$. In the Supplementary Materials we  prove that such states must be fully composed of dimer operators of the type $\sigma_i^\dagger- \sigma_j^\dagger$, see e.g. Eqs.~\eqref{eq:dimer},\eqref{eq:trimer}. This is a generalization of the single-excited dark dimer state $\sigma_i^\dagger|0\rangle- \sigma_j^\dagger|0\rangle$ in two atoms where the destructive interference suppresses photon emission~\cite{Brewer1996}. For a given set of numbers $N, m, k$, we can fully capture all the subradiant states by such a dimer basis. {Importantly, this dimer basis is not orthogonal, but it is still complete.}
The completeness of the basis also proves the  half-filling theorem for subradiant states:
the subradiant states only exist for $f \leq 1/2$. This   straightforwardly follows from the fact that 
the dimer basis can not be constructed for $f>1/2$ and generalizes the results in Ref.~\cite{Poshakinskiy2021dimer} for the two-level atom case.

\begin{figure}[t]
    \centering\includegraphics[width=0.48\textwidth]{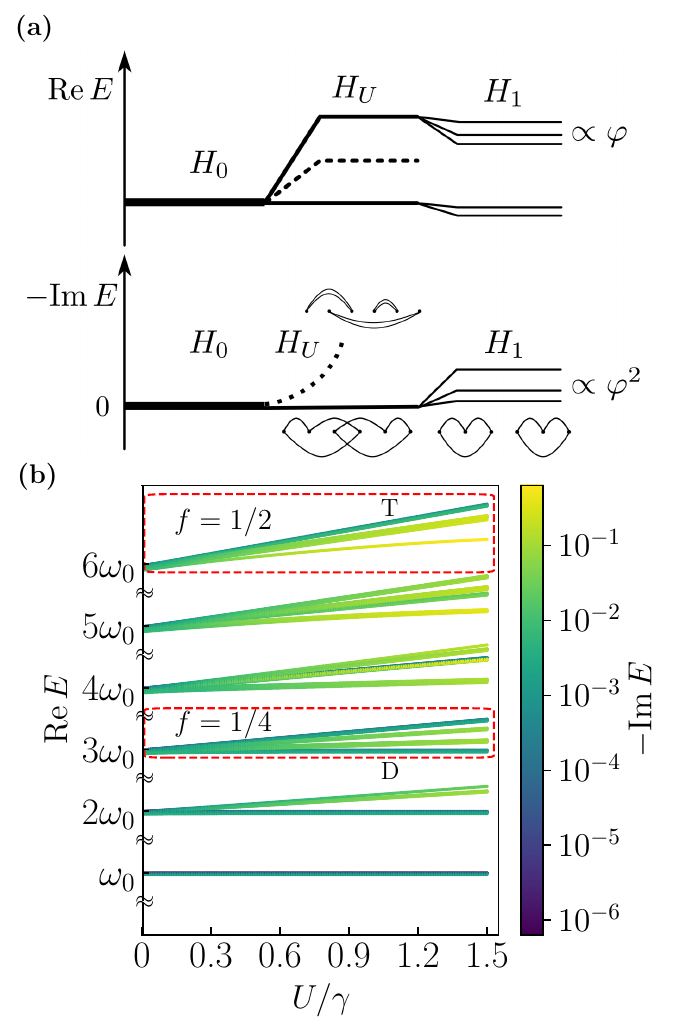}
 \caption{(a) A schematic figure of degenerate perturbations by the Hamiltonians~\eqref{eq:perturbation2}. 
{Small diagrams illustrate how the trimer basis states are determined by the degenerate perturbation in $H_U$ at $N=6,m=2,f=1/2$. }
(b) Dependence of energies for subradiant states in the 6-atom, 3-level array with phase $\varphi = 0.01$ on $U$. Red dotted boxes designate the states where $f=1/4,1/2$. The branches with dimer and trimer states are labeled by  ``D'' and ``T''.}
\label{fig:energyfan}
\end{figure}


Now we describe the degenerate perturbation theory in $H_U$ and $H_{1,2}$ to construct subradiant states, illustrated in Fig.~\ref{fig:energyfan}(a). {Both $H_U$ and $H_1$ lead to the splitting of the real part of the energy spectrum.} The anharmonic term $H_U$ also induces a nonzero decay rate for those states that are not eigenstates of $H_U$. These states would get mixed up with other non-subradiant states as the dashed line in Fig.~\ref{fig:energyfan}(a). 

{After the splitting by the anharmonicity potential}, 
{the remaining states are  shaped by the degenerate perturbation in $H_{1,2}$.} Their imaginary parts of energies are contributed by the first-order perturbation in $H_2$ 
and the second-order perturbation in $H_1$. {The real parts  are contributed by the expectation value of $H_U$, which is an integer multiple of $U$, and the first order perturbation of $H_1\propto-\varphi$.}

Figure~\ref{fig:energyfan}(b) illustrates this perturbation process by tracking the energy dependence of subradiant states on $U$ in a six-atom array with three levels per atom.
When the number of excitations is $k=1$, the states do not feel the effect of anharmonicity $H_U$ and  remain subradiant. When $k=2$, the subradiant states are split into three branches with different real parts of energies, $\Re E \approx 0, \frac{2}{3} U, \frac{5}{6} U$. The slopes can be analytically calculated when performing the degenerate perturbation. Only the states in the lowest branch are eigenstates of $H_U$ and stay subradiant. We define such a branch as a subradiant branch. On the other hand, the states in the upper two branches are not eigenstates of $H_U$ and thus become bright as $U$ is turned on. {Following the same procedure,} when $k=3$, there are four branches with $\Re E \approx 0,\frac{1}{3}U, \frac{5}{7}U , U$. Only the lowest and highest branches with integer slopes are subradiant. 
When $k=4$, there are five branches with $\Re E \approx 0.319U, U, U, 1.306U, 1.429U$. 
One of the branches with $\Re E \approx U$ is subradiant while the other one is not. When $k=5$, there are five branches, $\Re E \approx \frac{13}{18}U, \frac{10}{9}U, \frac{4}{3}U, \frac{5}{3}U$,
and none of them are subradiant. When $k=6$, there are three: $\Re E \approx \frac{13}{12}U, \frac{28}{17}U, 2U$, the highest branch are subradiant.

Among all the subradiant states above, we find that the dimer state at $f = 1/4$ is the most subradiant state in the branch $\Re E \approx 0$, and the trimer state at $f=1/2$ is the most subradiant state in the branch $\Re E \approx 2U$ as labeled in Fig.~\ref{fig:energyfan}(b). 
The intuitive explanation is that these multimer states, by construction, minimize the distances between the excitations and, as a result, minimize the expectation values of the Hamiltonians $H_1$ and $H_2$ in {Eq.}~\eqref{eq:perturbation2} providing the radiative decay. 
 {We note that
the half-filling requirement for an $m+1$-level atom means that each atom in the diagram of type Fig.~\ref{fig:1}(b,c) is connected to exactly $m$ neighbors. 

 {In order to further illustrate the formation of the trimer state in Fig.~\ref{fig:energyfan}(b), we introduce the concept of trimer basis. The difference between the full trimer basis and the particular trimer state in Eq.~\eqref{eq:trimer} is that the index 1,2,3 in Eq.~\eqref{eq:trimer} can be any other number set. Trimer basis states, as a subset  of dimer basis, are eigenstates of $H_U$. They remain subradiant while other states, like those involving products with more than one dimer operator $(\sigma_i^\dag-\sigma_j^\dag)^2$, get mixed up, as shown in Fig.~\ref{fig:energyfan}(a). After degenerate perturbation of $H_1$, the trimer state Eq.~\eqref{eq:trimer} wins as the most subradiant because of  its minimum expectation values of $ H_{1,2}$ among all the five trimer basis {states}. {Generally, at half-filling, only multimer basis states would stay subradiant after the degenerate perturbation of $H_U$. There is at most one subradiant branch. Therefore, if a multimer state exists at half-filling, it is the most subradiant state. }
}

For comparison, we also consider in the Supplementary Materials {another model of multilevel atoms that have $m+1$ equidistant levels}. In this model, there is no $H_U$ anharmonic term; hence, there is no multimer state for $f<1/2$. Instead,  approximate higher-order dimer states of type  $\psi_D=(\sigma_1^\dag-\sigma_2^\dagger)^m (\sigma_3^\dagger-\sigma_4^\dagger)^m|0\rangle$  become possible.  {Based on the intuitions above, such states with nearest-neighbor dimer couplings could be expected to be the most subradiant states at half-filling. However, we show in Supplementary Materials that such an expectation doesn't always hold.}

{\it Summary.}
Our findings provide a simple recipe for constructing {multiple-excited} subradiant states in {multilevel} atomic arrays. We show how the destructive interference and anharmonic level structure {favor} the multimer {product} states. We have tested this recipe in a one-dimensional array of multilevel atoms coupled to a waveguide. It is still an open question whether this intuition applies to two- or three-dimensional setups, which are now also emerging~\cite{DiLiberto2024} or to the waveguides with a more complex photon dispersion law~\cite{Zhang_2023}.



\bibliography{multimer}

\setcounter{figure}{0}
\setcounter{section}{0}
\setcounter{equation}{0}
\renewcommand{\thefigure}{S\arabic{figure}}
\renewcommand{\thesection}{S\Roman{section}}
\renewcommand{\thesection}{S\arabic{section}}
\renewcommand{\theequation}{S\arabic{equation}}

\newpage\clearpage
\section*{Supplementary materials}

\subsection{Subradiant dimer basis}

Here, we prove that subradiant states must be fully composed of dimer operators of the type $\sigma_i^\dagger- \sigma_j^\dagger$ at $\varphi \to 0$. First of all, we want to construct a mapping between the quantum raising operators and polynomials.

Any eigenstate with $k$ excitations  can always be written as 
\begin{equation}
    |\psi^{(k)}\rangle = \sum_{n_1 n_2\dots n_k=1}^N \psi_{n_1 n_2 \dots n_k} \sigma_{n_1}^\dagger \sigma_{n_2}^\dagger \dots \sigma_{n_k}^\dagger |0\rangle 
\end{equation}
where the tensor turns to zero if any of the $m+1$ indices coincide due to the limit of levels. Here, we define a mapping from the states to the homogeneous polynomials $P(a_1, a_2, \dots, a_N)$ by turning the creation operators $\sigma^\dagger_i$ into $a_{i}$ and the tensor $\psi$ into coefficients.

\begin{equation}
    |\psi^{(k)}\rangle \to P = \sum_{n_1 n_2\dots n_k=1}^N \psi_{n_1 n_2 \dots n_k} a_{n_1} a_{n_2}  \dots a_{n_k}\:.
\end{equation}

We also notice that the annihilation operator on a state $\sigma_i|\psi^(k)\rangle$ can be mapped to a derivative to the polynomial $\partial_{a_i} P$.

We are looking for the eigenstates of the Hamiltonian $H_0$ in Eq.~\eqref{eq:perturbation2} with zero eigenvalues
\begin{equation}
    H_0 |e\rangle = 0\:.
\end{equation}
We note  that $H_0$ can be written as 
\begin{equation}
    H_0 = \sum_{m,n=1}^N \sigma_m^\dagger \sigma_n = \left(\sum_{m=1}^N \sigma_m^\dagger\right) \left(\sum_{n=1}^N \sigma_n\right)\:.
\end{equation}
Therefore, a state $|e\rangle$ should fulfill the condition \begin{equation}
    (\sum_{n=1}^N \sigma_n) |e\rangle = 0
\end{equation}
so that $H_0 |e\rangle = 0$.
This equation  is equivalent to the following condition for the polynomials $P$
\begin{equation}
    \sum_{i=1}^N\partial_{a_i} P =0\:.
\end{equation}
Mathematically, it means that $P$ is a Lagrange polynomial, which can be written as $P(a_1-a_2, a_1-a_3,\dots, a_1-a_N)$. According to the mapping we defined, the subradiant states should be fully composed of dimer operators.

\subsection{Counting subradiant states when $U=0$.}

When there is no anharmonic term $H_U$, the number of subradiant states is equal to the number of linear independent dimer basis states. These states can be counted with straightforward but rather tedious combinatorics. Here, we present a rigorous formula to obtain the number of subradiant states with any given number of $N$, $m$, and $k$ as $D(N,m,k)$. Even when the anharmonic term exists, the formula provides the number of dimer basis states that will be useful for perturbation calculation.

First, we replace the operators $a_i$ in the previous section with $b_i = a_1-a_i$ for $2\leq i \leq N$. A dimer basis can be mapped to a homogeneous polynomial $P(a_1,a_2,\dots, a_N) \equiv P(b_2,b_3,\dots, b_N)$. So the number of linear independent dimer basis is equal to the number of all possible terms $b_2^{\nu_2} b_3^{\nu_3}\cdots b_N^{\nu_N}$ with a constraint $\sum_{i=2}^N \nu_i = k$ and $\text{deg}_{a_j}(P) \leq m, \forall j$. The number of excitations $k$ is thus the order of polynomial. We remind that the number of levels per atom is $m+1$.

Without the second constraint, the number of all possible terms is just the binomial coefficient 
 $   C_{N-2+k}^{N-2}$
that counts the number of ways to  put $k$ identical balls into $N-1$ different boxes.

Now we take into account the constraint $\text{deg}_{a_j}(P) \leq m, \forall j$. When $k\leq m$, this constraint is automatically satisfied. Therefore the number of subradiant states is just $D(N,m,k) = C_{N-2+k}^{N-2}$. However,  when $m+1 \leq k \leq 2m+1$,  we should subtract the number of terms that do not satisfy this constraint. There are two possible variants depending on the index of $a_j$. When $j\geq 2$, the constraint means that the terms like $b_2^{\nu_2} b_3^{\nu_3}\cdots b_N^{\nu_N}$ should obey $\nu_i \leq 2, \forall i\geq 2$. Here, the ``bad'' terms to be discarded can be constructed as putting $m+1$ balls into one box and distributing the rest into all the boxes. So, the number of discarded terms is
$
    C_{N-1}^1 C_{N-2+k-(m+1)}^{N-2}\:.
$
When $j=1$, the constraint means that we must make sure the degree of $a_1$ is no larger than $m$. We note, that $P$ satisfies $\sum_i \partial_{a_i} P =0$. Thus, we only need to set the coefficient of terms whose order of $a_1$ is $m+1$ to be zero, and  the terms with higher degrees of $a_1$ would be then zero automatically. As such, we  count the number of terms $a_1^{m+1}a_2^{\nu_2}\dots a_N^{\nu_N}$ with constraint $\sum_{i=2}^N \nu_i = k-m-1$
that is equal to 
\begin{equation}
    C_{N-2+k-(m+1)}^{N-2}\:.
    \label{eq:sa1}
\end{equation}
As a result,  when $m+1 \leq k \leq 2m+1$, the number of subradiant states is
$
    D(N,m,k) = C_{N-2+k}^{N-2} - N\cdot C_{N-2+k-(m+1)}^{N-2}\:.
$

The case when $2m+2 \leq k\leq 3m+2$ is even more complicated. The constraint on $b_2^{\nu_2} b_3^{\nu_3}\cdots b_N^{\nu_N}$, by the analogy of putting balls into boxes, demands that there should be no more than $m$ balls in $N-1$ boxes. There exist two kinds of bad terms that we have to exclude. (i) There are {at least} $2m+2$ balls in one box, or {at least} $m+1$ balls in two boxes. The number of balls in the rest boxes must be no more than $m$. The number of such bad terms is
\begin{equation}
    (C_{N-1}^1 + C_{N-1}^2)\cdot C_{N-2+k-(2m+2)}^{N-2}\:.
    \label{eq:sa2}
\end{equation}
(ii) There are {at least}   $m+1$ balls in one box. 
The remaining balls are then placed to make sure that the number of balls newly added into each box does not exceed $m$. The number of such bad terms is
\begin{equation}
    C_{N-1}^1 [D(N,m,k-m-1) + C_{N-2+k-(2m+2)}^{N-2}]\:.
    \label{eq:sa3}
\end{equation}
Here the term $C_{N-2+k-(2m+2)}^{N-2}$ comes from the constraint on $a_1$ in Eq.~\eqref{eq:sa1}. 

The constraint on $a_1$ introduces other types of terms $a_1^{m+1}a_2^{\nu_2}\dots a_N^{\nu_N}$ with a constraint $\sum_{i=2}^N \nu_i = k-m-1$ and $\nu_i \leq m$, since in Eq.~\eqref{eq:sa2} and Eq.~\eqref{eq:sa3} we have already excluded the terms with the degrees of $a_{2,3,\dots,N}$ larger than $m$:
\begin{equation}
    C_{N-2+k-(m+1)}^{N-2} - C_{N-1}^1 C_{N-2+k-(2m+2)}^{N-2}\:.
\end{equation}

With the calculation above, when $2m+2 \leq k\leq 3m+2$, the number of subradiant states is
\begin{equation}
    \begin{split}
        D(N,m,k) &= C_{N-2+k}^{N-2} - (C_{N-1}^1 + C_{N-1}^2) C_{N-2+k-(2m+2)}^{N-2} \\
    &- C_{N-1}^1 D(N,m,k-m-1) - C_{N-2+k-(m+1)}^{N-2} \:.
    \end{split}
\end{equation}

When $3m+3 \leq k \leq 4m+3$, following the similar procedure, we obtain the formula
\begin{equation}
    \begin{split}
        D(N,m,k) &= C_{N-2+k}^{N-2} - (C_{N-1}^1 + 2C_{N-1}^2 + C_{N-1}^3) \\ &\times C_{N-2+k-(3m+3)}^{N-2} \\
        &-(C_{N-1}^1 + C_{N-1}^2) D(N,m,k-2m-2) \\
        &- C_{N-1}^1  D(N,m,k-m-1)\\
        & - C_{N-2+k-(m+1)}^{N-2} \:.
    \end{split}
\end{equation}
One can further obtain the number of subradiant states for larger $k$. When $m=1$, the formulas above reduce to 
\begin{equation}
    C_N^k - C_N^{k-1}\:,
\end{equation}
which is the result in Ref.~\cite{Poshakinskiy2021dimer}.

There is another approach to calculate the number of subradiant states. The number of subradiant states at excitation $k$ is the difference of the number of eigenstates at $k$  and the number of eigenstates at $k-1$. While the number of eigenstates at $k$ is equal to the number of terms $a_1^{\nu_1}a_2^{\nu_2}\dots a_N^{\nu_N}$ with constraint $\sum_{i=1}^N \nu_i = k$ and $\nu_i \leq m, \forall i$. The idea is straightforward. The number of eigenstates at $k-1$ is the number of decay channels for eigenstates at $k$, {as illustrated in }Fig.~\ref{fig:Decay_channel}. If the number of decay channels is smaller than the number of eigenstates, some eigenstates do not decay and they are subradiant.

For example, at $N=4,m=2,k=4$, there are 
\begin{equation}
    a_1^2 a_2^2, \dots, a_1^2 a_2 a_3, \dots, a_1 a_2 a_3 a_4
\end{equation}
totally $6+12+1=19$ terms. At $k=3$, there are
\begin{equation}
    a_1^2 a_2, \dots, a_1 a_2 a_3, \dots
\end{equation}
totally $12+4=16$ terms. So the number of subradiant states at $k=4$ is $3$.

\begin{figure}[b]
\centering\includegraphics[width=0.4\textwidth]{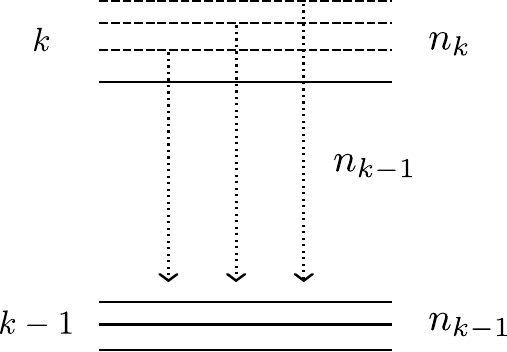}
\caption{Illustration of radiative decay channels between $n_k$ the states with the excitation number $k$ and $n_{k-1}$ states with $k-1$ excitations. {The number of subradiant states 
with excitation number $k$ is $n_k-n_{k-1}$}}
\label{fig:Decay_channel}
\end{figure}

We have checked that both approaches give the same result for the number of subradiant states as the numerical simulation {in Fig}.~\ref{fig:Number}.

\begin{figure}[b]
\centering\includegraphics[width=0.48\textwidth]{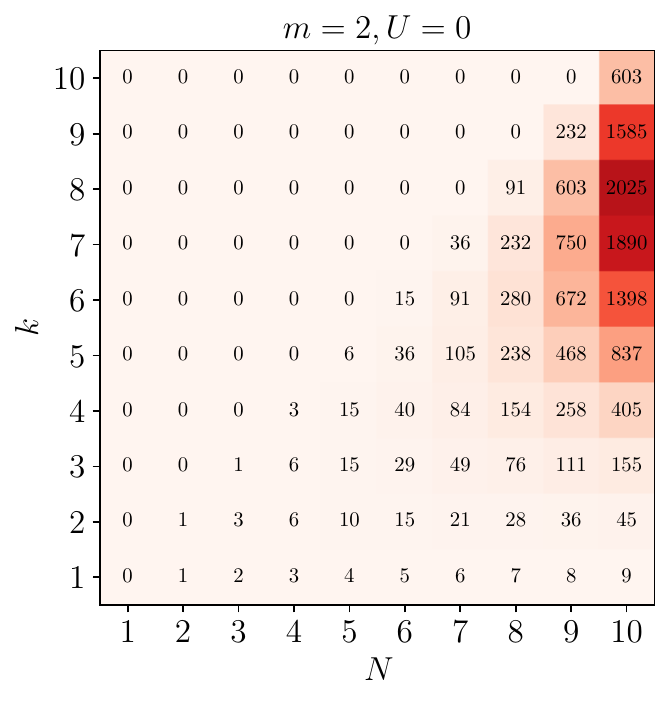}
\caption{{Numerical results for the} numbers of subradiant states at different excitations $k$ and atom number $N$ {in 3-level systems}. 
}
\label{fig:Number}
\end{figure}

\subsection{Energies of subradiant states}

Here, we present the details of the perturbation theory in $H_U$ and $H_{1,2}$ for subradiant states. The energies of subradiant states $|n_i\rangle$ at lowest order of $\varphi$ are
\begin{equation}
    E = k\omega_0 + \langle H_U \rangle - \Re E_1 \varphi - i(\Im E_1 + \Im E_2)\varphi^2\:.
    \label{eq:pertur_energy}
\end{equation}
The value of $\Re E_1$ is given by the first order perturbation in $H_1$,
\begin{equation}
    \Re E_1 = \langle n_i |H_1| n_i\rangle\:,
\end{equation}
and the value
$\Im E_2$ is given by the first order perturbation in $H_2$:
\begin{equation}
    \Im E_2 = \langle n_i |H_2|n_i\rangle /2\:.
\end{equation}
The factor of  $1/2$ here stems from Eq.~\eqref{eq:perturbation} in the main text, that is $H_I=\ldots-\rmi \varphi^2H_2/2$.

The imaginary part $\Im E_1$ is given by the second order perturbation in $H_1$. It involves all the other non-subradiant states $|e\rangle$ at the same excitation number since we assume that $\omega_0/\gamma_{\rm 1D} \gg 1$,
\begin{equation}
    \begin{split}
        &\Im E_1 =\\ &-\Im\left(\sum_{e} \frac{\langle n_i|H_1|e\rangle\langle e| H_1|n_i\rangle}{\langle n_i|H_U|n_i\rangle- \langle e|H_U|e\rangle + \rmi \langle e| H_0|e \rangle} \right)
    \end{split}\:.
    \label{eq:ImE2}
\end{equation}
For  $U=0$, this expression reduces to 
\begin{equation}
    \Im E_1 = \langle n_i|H_1 \sum_e \frac{|e\rangle\langle e|}{\langle e|H_0|e\rangle} H_1 |n_i\rangle\:.
    \label{eq:ImE1}
\end{equation}

Since we do not have an explicit equation to  analytically calculate all the eigenstates of $H_0$, the value $\Im E_1$ can only be obtained numerically. In numerical calculation of  the eigenstates of $H_0$, it is convenient to introduce small perturbation $\propto H_1$ to split the degeneracy.

\subsection{Multilevel atoms with equidistant levels. }

In the main text we focus on the atoms described by a Bose-Hubbard-type model, where the anharmonic term $H_U$ splits the states into different branches as shown in Fig.~\ref{fig:energyfan}. Here, we consider an atomic array with a finite number of equidistant levels per atom.

 Figure~\ref{fig:minS-f-U0} shows the fill-factor dependence of the minimum value of entanglement entropy for the equidistant level model. Comparing with  Fig.~\ref{fig:minS-f} in the main text, we find that when $U=0$, we find no multimer state  below half-filling. At half-filling, in addition to the trimer states at $N=6,9$ in Fig.~\ref{fig:minS-f-U0}(a), there appears an approximate higher-order dimer state at $N=8$ with nearly zero entanglement entropy:
\begin{equation}
    (\sigma_1^\dagger-\sigma_2^\dagger)^2(\sigma_3^\dagger-\sigma_4^\dagger)^2(\sigma_5^\dagger-\sigma_6^\dagger)^2(\sigma_7^\dagger-\sigma_8^\dagger)^2|0\rangle\:.
\end{equation}

In Fig.~\ref{fig:minS-f-U0}(b), at half-filling, we find another approximate higher-order dimer state at $N=8$ instead of the tetramer state.
\begin{equation}
    (\sigma_1^\dagger-\sigma_2^\dagger)^3(\sigma_3^\dagger-\sigma_4^\dagger)^3(\sigma_5^\dagger-\sigma_6^\dagger)^3(\sigma_7^\dagger-\sigma_8^\dagger)^3|0\rangle\:.
\end{equation}
At $N=6, m=3, f=1/2$,   {higher-order dimer states are also possible}. 

The fact that no multimer state survives below half-filling emphasizes the importance of the anharmonic term $H_U$. Without the degenerate perturbation of $H_U$ in Fig.~\ref{fig:energyfan}, these multimer states would get mixed up by the degenerate perturbation of $H_1$. 

{It is not clear yet why the multimer states still exist above half-filling, but this effect is out of the scope of the current study.}

\begin{figure}[t]
    \centering\includegraphics[width=0.49\textwidth]{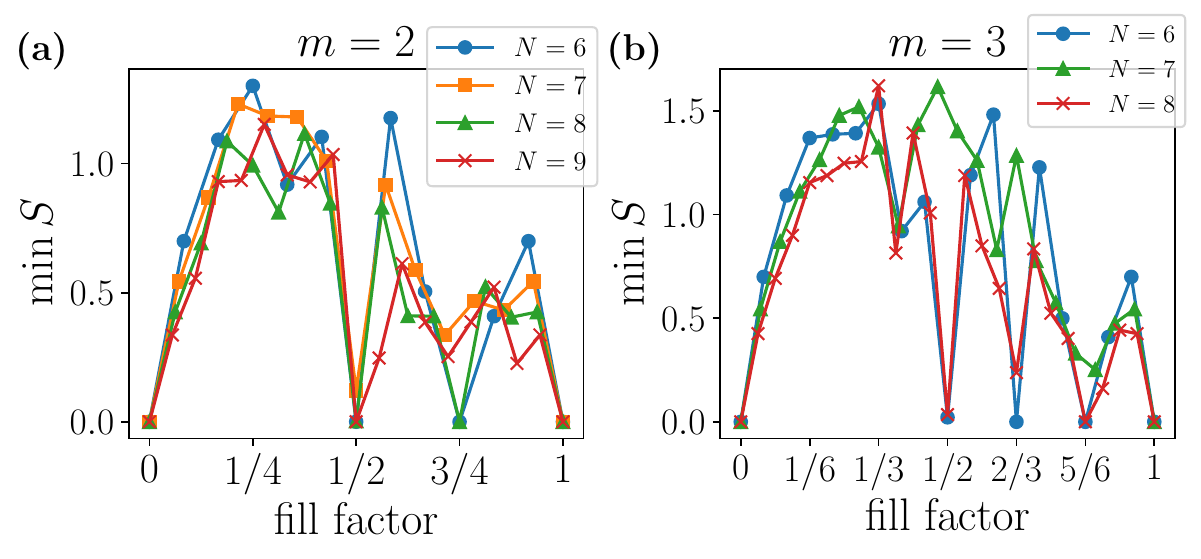}
\caption{Dependence of the minimum value of entanglement entropy on fill factor $f$ calculated at $\varphi = 0.001$  for (a) $m=2$ and (b) $m=3$.}
\label{fig:minS-f-U0}
\end{figure}
\begin{table*}[tbh!]
	\centering
	\resizebox{0.8\textwidth}{!}{
		\begin{tabular}{|c|c|c|c|c|c|c|c|c|c|c|}
			\hline
			$m$ & 2 & 3 & 4 & 5 & 6 & 7 & 8 & 9 & 10 \\
			\hline
			$\mathrm{Im}\, E_1$ & 0.38888 & 1.82 & 5.6977 & 15.5693 & 41.10887 & 110.144 & 306.0854 & 887.6915 & 2679.763 \\
			\hline
			$|\delta_0|^2$ & 1 & 3 & 5.5 & 8.125 & 10.6875 & 13.125 & 15.4375 & 17.6484375 & 19.78515625 \\
			\hline
			$\sum_e (\langle e|H_0|e\rangle)^{-1}$ & 6.304762 & 15.9961 & 34.78263 & 71.52311 & 147.5097 & 317.7914 & 733.2622 & 1827.4642 & 4894.2766 \\
			\hline
			$\alpha$ & 0.06168 & 0.03793 & 0.2978 & 0.02679 & 0.02608 & 0.02641 & 0.02704 & 0.02752 & 0.02767\\
			\hline
		\end{tabular}
	}
 \caption{Values of $\Im E_1$, $|\delta_0|^2$, $\sum_e (\langle e|H_0|e\rangle)^{-1}$ and the fitting coefficient $\alpha$ for different values of $m$ in 4-atom array systems.}
 \label{table1}
\end{table*}

In the absence of $H_U$, as shown in Fig.~\ref{fig:minS-f-U0}, there are higher-order dimer states at half-filling. We will now discuss when such states are exact eigenstates and when they are the most subradiant ones. The general expression of higher-order dimer states is
\begin{equation}
    |n_0\rangle = (\sigma_1^\dagger-\sigma_2^\dagger)^m (\sigma_3^\dagger-\sigma_4^\dagger)^m \cdots (\sigma^\dagger_{N-1}-\sigma_N^\dagger)^m |0\rangle\:.
    \label{eq:high_dimer}
\end{equation}

In a two-level atomic system, $m=1$, the dimer state is an eigenstate of $H_1$ with eigenvalue $\frac{N}{2}$. So it's exact. However, when $m>1$, $|n_0\rangle$ is not an eigenstate of $H_1$,
\begin{equation}
    H_1 |n_0\rangle = \frac{mN}{2} |n_0\rangle + |\delta_0\rangle \:.
\end{equation}

When $N=4$, the expression for the exact term is $|\delta_0\rangle = -m(\sigma_3^\dagger + \sigma_4^\dagger)(\sigma_1^\dagger-\sigma_2^\dagger)^{m-1} (\sigma_3^\dag-\sigma_4^\dagger)^m + m(\sigma_1^\dagger + \sigma_2^\dagger)(\sigma_1^\dagger-\sigma_2^\dagger)(\sigma_3^\dagger-\sigma_4^\dagger)^m|0\rangle$. We remind that $(\sigma^\dagger)^m |0\rangle = 0$. We find that $|\delta_0\rangle$ is perpendicular to any dimer basis $|n_i\rangle$ because of the symmetry: after swapping $\sigma_1^\dagger$ and $\sigma_2^\dagger$, $\sigma_3^\dagger$ and $\sigma_4^\dagger$, $|n_i\rangle$ is still $|n_i\rangle$ but $|\delta_0\rangle$ becomes $-|\delta_0\rangle$. So for $N=4$, higher-order dimer states are still exact eigenstates.

When $N>4$, $|n_0\rangle$ gets mixed up with the other states. However, the degree of mixing is small. {The qualitative reason is that even though  $|n_0\rangle$ is not an exact eigenstate of $H_1$, it still has the lowest possible expectation value of $\langle H_1\rangle$ among all the dimer basis states. This is because this state involves only the dimers with adjacent sites.} As a result, when solving the degenerate perturbation equation $\mathrm{det}[\langle n_i|H_1|n_j\rangle - x\langle n_i|n_j\rangle] = 0$, one would get an eigenvector close to $(1,0,\dots,0)$ with the eigenvalue close to $\frac{mN}{2}$. After the perturbation, one would get the eigenstate $|e_0\rangle \approx |n_0\rangle$.

We will now demonstrate that the higher-order dimer state is the most subradiant one only for small values of $m$ and $N$. 
By construction, $|n_0\rangle$ has also the minimum expectation value of $\langle H_2\rangle = mN/2$. Then according to Eq.~\eqref{eq:ImE2}, the value $\Im E_2 =mN/4$. This makes $|n_0\rangle$  a good candidate for the most subradiant state. However, another contribution to the imaginary part of the energy, $\Im E_1$ grows exponentially with $m$. As a result, it turns out that for large of $m$, the state $|n_0\rangle$ becomes surpassed by other subtadiant states. To demonstrate this, we take the case $N=4$ as an example. After normalization of Eq.~\eqref{eq:high_dimer} with $\langle n_0|n_0\rangle = (2^m m!)^2$, we find that $\langle \delta_0|\delta_0\rangle = 2m \frac{2^m -m-1}{2^m}$.

\begin{figure}[h!]
\centering\includegraphics[width=0.45\textwidth]{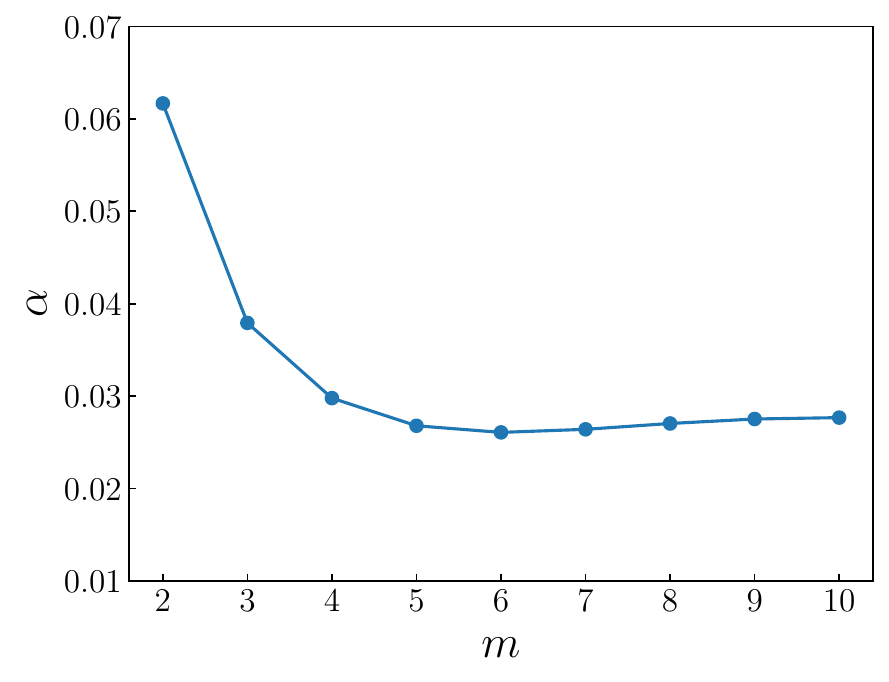}
\caption{Dependence of $\alpha$ on $m$}\label{fig:alpha}
\end{figure}

Since $|n_0\rangle$ is orthogonal to non-subradiant states, we take an approximation of Eq.~\eqref{eq:ImE1},
\begin{equation}
    \begin{split}
        \Im E_1 &= \langle n_0|H_1 \sum_e \frac{|e\rangle\langle e|}{\langle e|H_0|e\rangle} H_1 |n_0\rangle \\
        &\approx \alpha \langle \delta_0|\delta_0\rangle (\sum_e \frac{1}{\langle e|H_0|e\rangle}) \:.
    \end{split}
\end{equation}

{One can see in Table~\ref{table1} and Fig.~\ref{fig:alpha}, that when $m\geq 4$ this approximation works  well.}


Our numerical finding is that  $\ln|\Im E_1| \propto m$. And for $N=4$ and $m=5$ the state $|e_0\rangle$ becomes the second most subradiant state. For $N \geq 6$, when $m=3$, $|e_0\rangle$ is not the most subradiant. For $m=2$, the maximum value of $N$ that we were able to consider numerically on a workstation with 150 Gb of RAM was 10. We have verified that the state  $|e_0\rangle$  stays the most subradiant until this value of $N$.

\subsection{Shadows of the multimer states 
}

In the main text, we demonstrate the existence of the multimer states that are direct products of dimers, trimers, tetramers, etc. In the previous section (dimer basis) we show that the physical reasons behind the multimerization effect is highly relevant with subradiance. This intuition can be supported by some states that look like shadows of the multimer states. These states appear when the number of atoms  is not an integer multiple of $3$,$4$, etc., and does  not meet the requirements for multimerization. They are not direct product states but are related to the multimer states. 

For example, in a 5-atom array with 3 levels per atom, when $\varphi \to 0$, the most subradiant state with 4 excitations is
\begin{equation}
    \begin{split}
        &0.213(\sigma_1^\dagger-\sigma_2^\dagger) (\sigma_3^\dagger-\sigma_4^\dagger) (\sigma_4^\dagger- \sigma_5^\dagger) (\sigma_5^\dagger - \sigma_3^\dagger)|0\rangle\\
        -& 0.213(\sigma_1^\dagger-\sigma_2^\dagger)(\sigma_2^\dagger-\sigma_3^\dagger)(\sigma_3^\dagger - \sigma_1^\dagger)(\sigma_4^\dagger-\sigma_5^\dagger) |0\rangle\\
        +& 0.041 (\sigma_2^\dagger-\sigma_3^\dagger)(\sigma_1^\dagger - \sigma_4^\dagger)(\sigma_4^\dagger - \sigma_5^\dagger)(\sigma_5^\dagger - \sigma_1^\dagger) |0\rangle\\
        -& 0.041(\sigma_3^\dagger-\sigma_4^\dagger) (\sigma_1^\dagger -\sigma_2^\dagger) (\sigma_2^\dagger-\sigma_5^\dagger)(\sigma_5^\dagger-\sigma_1^\dagger) |0\rangle
    \end{split}
\end{equation}
One can  {neglect the last two terms} and 
{approximate} this state as
\begin{equation}
    |{\rm dimer}\rangle \otimes |{\rm trimer}\rangle - |{\rm trimer}\rangle \otimes |{\rm dimer}\rangle
\end{equation}
While such states are not direct product state and their entanglement entropy is nonzero, they further demonstrate the tendency of the system to form multimer states.

\end{document}